\documentclass[aip,reprint,twocolumn,showpacs,prb,amssymb,amsmath,floatfix]{revtex4}

\usepackage{graphicx}
\usepackage{dcolumn}
\usepackage{bm}
\usepackage{color}
\begin{document}

\title{Ba$_{2}$TeO as an optoelectronic material: First-principles study}

\author{Jifeng Sun$^{1,2,3}$}
\author{Hongliang Shi$^{1}$}
\author{Mao-Hua Du$^{1}$}
\author{Theo Siegrist$^{2,3}$}
\author{David J. Singh$^{1}$}
 \affiliation{$^1$Materials Science and Technology Division, Oak Ridge National Laboratory, Oak Ridge, TN 37831, USA}
 \affiliation{$^2$Florida Agricultural $\&$ Mechanical University-Florida State University, College of Engineering, Department of Chemical $\&$ Biomedical Engineering, 2525 Pottsdamer St., Tallahassee, FL 32310}
  \affiliation{$^3$National High Magnetic Field Laboratory, 1800 E. Paul Dirac Dr., Tallahassee, FL, 32310}
\date{\today}

\begin{abstract}
The band structure, optical and defects properties of Ba$_{2}$TeO are systematically investigated using density functional theory with a view to understanding its potential as an optoelectronic or transparent conducting material. Ba$_{2}$TeO crystallizes with tetragonal structure (space group P4/nmm) and with a 2.93 eV optical band gap \citep{r14}. We find relatively modest band masses for both electrons and holes suggesting applications. Optical properties show a infrared-red absorption when doped. This could potentially be useful for combining wavelength filtering and transparent conducting functions. Furthermore, our defect calculations show that Ba$_{2}$TeO is intrinsically p-type conducting under Ba-poor condition. However, the spontaneous formation of the donor defects may constrain the p-type transport properties and would need to be addressed to enable applications.     
\end{abstract}

\pacs{Computational Physics, Condensed Matter Physics}
\maketitle

\section{\label{sec:level1}INTRODUCTION}
Oxychalcogenides have been extensively investigated as these materials exhibit promising properties in a variety of areas including transparent conductors, semiconductors, optoelectronics as well as thermoelectrics.\citep{r7,r8,r30,r31,r32,r33,r35} The semiconducting oxychalcogenides (e.g., LnCuO$Ch$, $Ch$ = chalcogenides) with layered structures have interesting optoelectronic properties with wide band gaps and many of them are p-type transparent conductors.\citep{r29,r34} The high hole conductivity stems from the more dispersive chalcogen p states that make up the valence band maximum instead of the O 2p states in traditional oxides. Multifunctional materials, including materials that combine transparent conducting behavior with wavelength filtering may find applications in solar and display technologies.\citep{r28}     

Barium chalcogenides (BaX, X=O, S, Se, Te) are all wide band gap semiconductors with band gaps ranging from 3.08 eV to 4.10 eV.\cite{r6,r5} Incorporatinon of additional anions with more dispersive p bands into wide-band-gap oxides (e.g. ZrOS, HfOS) could be an effective mechanism in designing new p-type transparent conducting materials (TCMs).\cite{r15} Here we study the newly found ternary barium oxytelluride, Ba$_{2}$TeO, which consists of alternating BaTe and BaO layers and has an experimental optical band gap value of 2.93 eV.\cite{r14} We present first-principles calculations of electronic and optical properties for doped material and studies of native defects in Ba$_{2}$TeO. We found relatively small effective masses for both hole and electron carriers. Our optical calculations indicate potential as an energy filtering into TCM if suitable doping can be achieved. However, based on the energetics of the native defects, achieving this is likely to be difficult.

\section{\label{sec:level2}Computational methods}
Our band structures and optical properties were computed using the full-potential linearized augmented plane-wave and local orbitals (FP-LAPW+lo) method \cite{r16} as implemented in the WIEN2K code.\cite{r17} We did calculations with both the Perdew-Burke-Ernzerhof (PBE) \cite{r24} functional and the modified Becke-Johnson potential (mBJ) functional.\cite{r18} The latter functional generally predicts improved band gaps for semiconductors and insulators compared to traditional functionals.\cite{r18,r19,r20} The LAPW sphere radii were 2.6 bohr for both Ba and Te, and 2.0 bohr for O. The cut-off parameter for the plane wave basis was $R_{min}K_{max}=7$. For the perfect crystal, experimental lattice constants were used and the atomic coordinates were fully optimized by minimizing the forces down to 1 mRy/a.u.. The convergence was tested and an 8$\times$8$\times$8 k-mesh with 75 k-points in the irreducible Brillouin zone was employed for the self-consistent calculation and a denser k-mesh of 16$\times$16$\times$16 for the optical properties. The calculations for doped Ba$_{2}$TeO were performed within the virtual crystal approximation. This average-potential approximation is often suitable for semiconductors especially when there is a electropositive site (Ba in the present case) where doping is performed.\cite{r28} Both n-type and p-type doping were considered on the Ba sites, with 0.1 electrons (holes)/unit cell and 0.2 electrons (holes)/unit cell, and the corresponding concentrations are 4$\times10^{20}$ cm$^{-3}$ and 8$\times10^{20}$ cm$^{-3}$, respectively. Spin-orbit coupling (SOC) was included.

The formation energies of possible intrinsic defects were calculated using the projector augmented-wave (PAW) method \cite{r21} implemented in the VASP code.\cite{r22} A 144 atoms supercell constructed from the primitive unit cell was used. A 2$\times$2$\times$2 k-mesh was generated within the Monkhorst-Pack scheme.\cite{r23} The PBE \cite{r24} type exchange-correlation functional was employed. Because of the underestimation of the band gap from PBE, we also performed the self consistent HSE calculations \cite{r25} to correct the band gap. Within the HSE calculations, the valence band maximum (VBM) and conduction band minimum (CBM) shifted down by 0.91 eV (VBM) and up by 0.34 eV (CBM), respectively.

The formation of the defects can be considered as an exchange process between the host atoms/electrons and other atomic and electronic reservoirs. Therefore, the total energy changes due to the particle changes. For a system consisting of defects $\alpha$ that are ionized to the charge state q, the total formation energy can be computed as \cite{r26}

\begin{eqnarray}
\Delta H_{\alpha,q}(E_{F},\mu)=&E_{\alpha,q}-E_{host}+\underset{\alpha} {\sum}n_{\alpha}\mu_{\alpha}\nonumber\\
&+q(E_{VBM}+E_{F}).
\label{eq1}
\end{eqnarray}

Where $E_{\alpha,q}$ is the total energy of the supercell with defect $\alpha$ in charge state q, and $E_{host}$ is just the total energy of the pure host supercell. $n_{\alpha}$ is the number of removed defect atoms $\alpha$: $n_{\rm{O}}=1$ for an oxygen vacancy. $\mu_{\alpha}$ is the chemical potential of species $\alpha$ which can be further expended to $\mu_{\alpha}=\Delta\mu_{\alpha}+\mu_{\alpha}^{elemental}$. Here $\Delta\mu_{\alpha}$ is the chemical potential of the $\alpha$th atom relative to it's elemental phase. $E_{VBM}$ and $E_{F}$ are the energy at VBM and the Fermi energy of the perfect system, respectively. 

In general, $\mu_{\alpha}$ is not a free variable, and in our system, $\mu_{\rm{Ba}}$, $\mu_{\rm{Te}}$, and $\mu_{\rm{O}}$ are limited by the synthesis conditions to precipitate the pure Ba$_{2}$TeO crystals. First, $\mu_{\alpha}$ should be smaller than $\mu_{\alpha}^{elemental}$ to prevent the formation of the elemental phase. This means:

\begin{equation}
\Delta\mu_{\rm{Ba}}\leqslant 0,\quad\Delta\mu_{\rm{Te}}\leqslant0,\quad\Delta\mu_{\rm{O}}\leqslant0;
\end{equation}
 \begin{normalsize}
 
 \end{normalsize}

The point that $\Delta\mu_{\alpha}=0$ represents the rich condition of chemical species $\alpha$. Second, competing secondary phases need to be avoided by satisfying the following conditions:

\begin{equation}
\Delta\mu_{\rm{Ba}}+3\Delta\mu_{\rm{Te}}\leqslant\Delta H_{f}(\rm{BaTe_{3}})
\end{equation}

\begin{equation}
\Delta\mu_{\rm{Ba}}+2\Delta\mu_{\rm{Te}}\leqslant\Delta H_{f}(\rm{BaTe_{2}})
\end{equation}

\begin{equation}
\Delta\mu_{\rm{Ba}}+\Delta\mu_{\rm{Te}}\leqslant\Delta H_{f}(\rm{BaTe})
\end{equation}

\begin{equation}
\Delta\mu_{\rm{Ba}}+\Delta\mu_{\rm{O}}\leqslant\Delta H_{f}(\rm{BaO})
\end{equation}

and

\begin{equation}
\Delta\mu_{\rm{Ba}}+2\Delta\mu_{\rm{O}}\leqslant\Delta H_{f}(\rm{BaO_{2}})
\end{equation}

\begin{figure}
\includegraphics*[height=7cm,keepaspectratio]{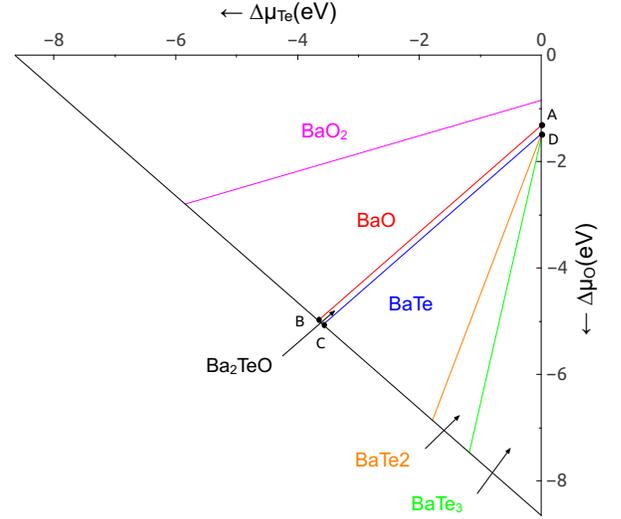}
\caption{\label{fig1}Chemical potential diagram for Ba-Te-O ternary system (Ba$_{2}$TeO) and it's other competing phases, including elemental phases Ba, Te and O, and binary phases BaO$_{2}$, BaO, BaTe, BaTe$_{2}$ and BaTe$_{3}$.}
\end{figure}

Where the formation energy of the five secondary phases were calculated to be: $\Delta H_{f}(\rm{BaTe_{3}})=-3.54$ eV, $\Delta H_{f}(\rm{BaTe_{2}})=-3.57$ eV, $\Delta H_{f}(\rm{BaTe})=-3.58$ eV, $\Delta H_{f}(\rm{BaO})=-4.98$ eV and $\Delta H_{f}(\rm{BaO_{2}})=-5.59$ eV. We have considered BaO as a competing phase in the synthesis of Ba$_{2}$TeO. Thus, Ba$_{2}$TeO is thermodynamically stable only in the narrow region between BaTe and BaO, as shown in Fig. \ref{fig1}. We did not consider reduced phases that could occur in oxygen deficient conditions.

The defect transition energy level, $\epsilon_{\alpha}$(q/q$^{'}$) is defined as the energy where the formation energy of $\alpha$ with charge state q equals that of charge state q$^{'}$, i.e., $\Delta H_{\alpha,q}=\Delta H_{\alpha,q^{'}}$. Using Eq. \ref{eq1}, the transition energy can be calculated as:

\begin{equation}
\epsilon_{\alpha}(q/q^{'})=[E_{\alpha,q}-E_{\alpha,q^{'}}]/(q^{'}-q)
\label{eq8}
\end{equation}

Both the band filling correction and the potential alignment correction were performed for the charged defects. The PBE band gap error was also corrected by the HSE calculation. The common correction details can be found for example in Ref. \citep{r27}. Importantly, in the HSE calculations we chose the mixing parameter to reproduce the experimental band gap.

\section{\label{sec:level3}RESULTS AND DISCUSSIONS}

\subsection{\label{sec:1}Band structures and effective masses}

\begin{figure}
\includegraphics*[height=8cm,keepaspectratio]{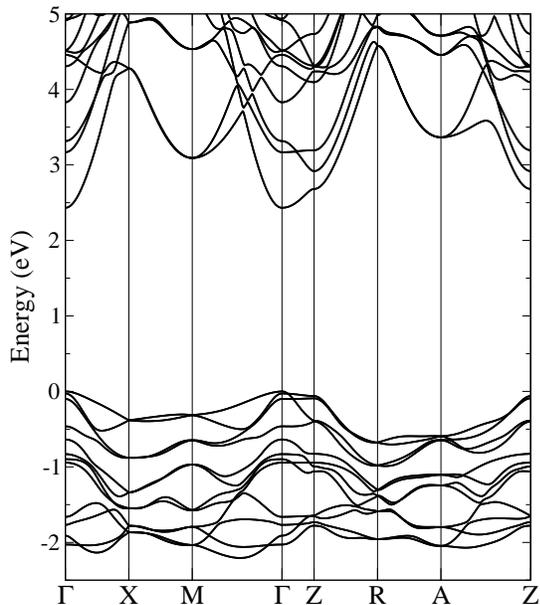}
\caption{\label{fig2}Calculated band structures of Ba$_{2}$TeO using mBJ+SO. The energy zero is set at the valence maximum.}
\end{figure}

Ba$_{2}$TeO was found to be a semiconductor with a band-gap about 2.93 eV. \citep{r14} The band structure calculated in mBJ is shown in Fig. \ref{fig2}, with spin-orbit included. There is a direct band gap at $\Gamma$ point with a value of 2.43 eV. PBE calculations, which underestimate gaps in general yield a 1.67 eV gap also direct at $\Gamma$. The valence bands are mainly Te 4p and O 2p characters with Te 4p at the VBM. Electron and hole effective masses along $\Gamma$-X and $\Gamma$-Z direction for both VBM and CBM are given in Table \ref{tab:table1}. The calculated energy separations between the VBM and the band next to VBM (VBM-1) are about 28 meV at $\Gamma$ point at the mBJ level. Since this energy difference is comparable to kT at room temperature, the light holes on the VBM-1 can potentially be populated once the holes are introduced into the valence bands at high temperatures or by heavy doping. As can be seen from Table \ref{tab:table1}, both the hole and electron effective masses are small enough that the material could have reasonable transport properties for both n- and p-type.  Furthermore, although the crystal structure has a layered appearance considering the top valance band, the band along $\Gamma$-Z is more dispersive than along $\Gamma$-X, which is also apparent from the hole effective mass listed in Table I. Therefore, compared to layered p-type compounds like LaCuOS and LaCuOSe, Ba$_{2}$TeO may have better hole conductivity along the c axis due to the smaller hole effective mass.

\begin{table}
\caption{\label{tab:table1} Calculated hole and electron effective masses of CBM and top two valence bands, VBM and VBM-1,  along $\Gamma$-X and $\Gamma$-Z in Ba$_{2}$TeO using mBJ functional with spin-orbit coupling. All effective masses given in m$_e$.}

\begin{ruledtabular}
\begin{tabular}{cccc}
       &  $\Gamma$-X    &     $\Gamma$-Z \\
\hline \\
CBM    & 0.29      &     0.96 \\
VBM    & 1.06      &     0.29 \\
VBM-1  & 0.28      &     3.24 \\
\end{tabular}
\end{ruledtabular}
\end{table}

\subsection{\label{sec:2}Optical properties}

\begin{figure}
\includegraphics*[height=5.5cm,keepaspectratio]{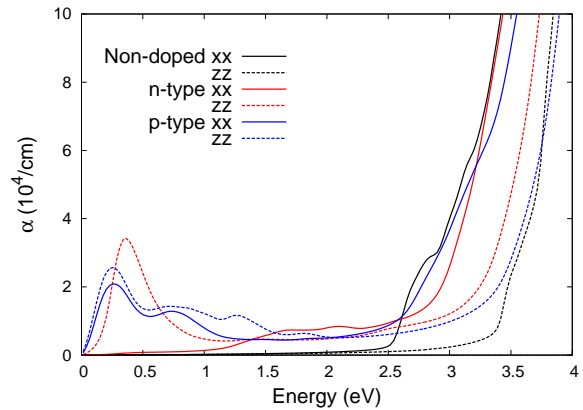}
\caption{\label{fig3}Calculated absorption spectra (no Drude) for doped and un-doped Ba$_{2}$TeO with doping concentration of 0.2 holes/electrons per unit cell (8$\times10^{20}$ cm$^{-3}$). The solid lines are for the in-plane direction and dashed lines represent the out-of-plane direction.}
\end{figure}

The calculated absorption coefficients for the doped Ba$_{2}$TeO with a carrier concentration of 8$\times10^{20}$ cm$^{-3}$, together with the un-doped data are given in Fig. \ref{fig3}, showing both in-plane (solid lines) and out-of-plane (dash lines) directions. Substantial absorption can be seen in the infrared and red energy range (below 2 eV). In particular, it shows strong absorption in c-axis direction, but no absorption for the polarization in the plane for n-type doping until  about 1 eV. On the other hand, p-type doping yields similar moderate absorption in both directions.

\begin{figure}
\includegraphics*[height=5.7cm,keepaspectratio]{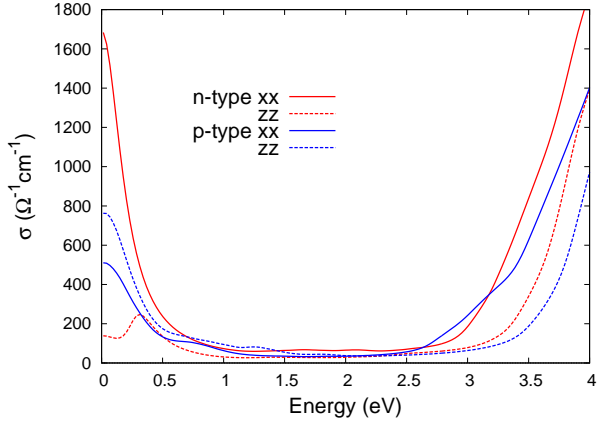}
\caption{\label{fig4}Optical conductivities with Drude contributions for both p-type and n-type doped Ba$_{2}$TeO at 0.2 holes/electrons per unit cell (8$\times10^{20}$ cm$^{-3}$). The solid lines are for the in-plane direction and dashed lines represent the out-of-plane direction.}
\end{figure}

Fig. \ref{fig4}. depicts the optical conductivity of the doped compounds, in which the Drude contribution is included, with an assumed broadening parameter $\gamma$ = 0.2. The important feature is that the n-type high optical conductivity caused by the Drude contribution in the plane is much stronger than the c-axis. This stems from the larger intra-band transitions along $\Gamma$-X comparing with $\Gamma$-Z directions in the corresponding conduction bands since the mobility of the electron is higher in the plane. In the p-type compounds, no dramatic difference was found within the intra-band contribution, but with c-axis having higher values.

\subsection{\label{sec:3} Defect formation energies and transition levels}
Native defects that act as acceptors or donors are crucial in the understanding of the hole generation and compensation process for p-type TCMs. We studied three important types of native point defects: Ba, Te and O vacancies ( V$_{\rm{Ba}}$, V$_{\rm{Te}}$, V$_{\rm{O}}$). In Ba$_{2}$TeO, the crystal structure can be considered as the combination of BaTe layers (containing Ba1) and BaO layers (containing Ba2).

\begin{table}
\caption{\label{tab:table2} Chemical potentials of the component atoms of Ba$_{2}$TeO at points A-D in Figure \ref{fig1}.}
\begin{ruledtabular}
\begin{tabular}{cccccc}
Point   & $\Delta\mu_{Ba}$(eV)     & $\Delta\mu_{Te}$(eV)  & $\Delta\mu_{O}$(eV) \\
\hline \\
A    & -3.66      &     0     &       -1.32      \\
B    &  0         &   -3.66   &       -4.98      \\
C    &  0         &   -3.58   &       -5.06      \\
D    & -3.58      &     0     &       -1.48      \\
\end{tabular}
\end{ruledtabular}
\end{table}

\begin{figure}
\includegraphics*[height=12cm,keepaspectratio]{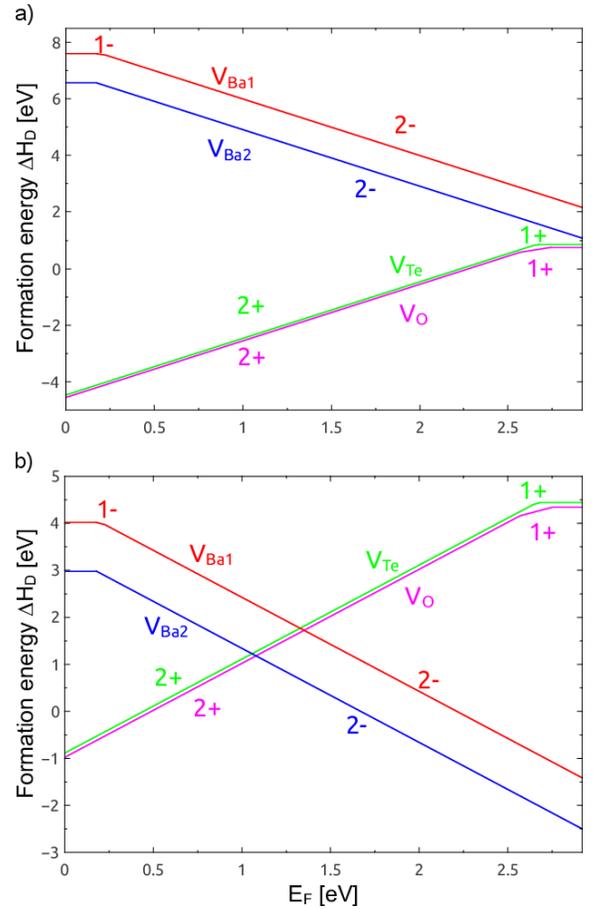}
\caption{\label{fig5}Calculated formation energy of defects as a function of the Fermi energy E$_{f}$, with (a) represents the point C (Ba-rich) and (b) the point D (Ba-poor) in Fig. \ref{fig1}, respectively.}
\end{figure}

Our calculated formation energies are plotted in Fig. \ref{fig5} (a) and (b) for these possible intrinsic defects. Two chemical potential conditions are considered here, the point C which corresponds to the Ba-rich condition, and the point D which corresponds to the Ba-poor condition. The chemical potentials corresponding to the phase diagram (Fig. \ref{fig1}) from point A to D are listed in Table \ref{tab:table2}.

As can be seen, it is clear that barium vacancies on either site are shallow acceptors with V$_{\rm{Ba2}}$ the dominant acceptor and the anion vacancies are shallow donors where V$_{\rm{O}}$ has the lowest formation energy. Particularly, under the Ba-rich condition, the formation energy of ionized V$_{\rm{Te}}$ and V$_{\rm{O}}$ with q = +2 are so low that they become negative starting from the CBM. However, the formation energies of barium vacancies are high enough that no crossing point could occur between anion vacancies and cation vacancies. The negative formation energy of V$_{\rm{Te}}$ and V$_{\rm{O}}$ also implies that the charged anion vacancies can form spontaneously and act as hole killers when the Fermi level is close to CBM. Therefore the free hole concentration would be limited under this condition.

\begin{figure}
\includegraphics*[height=6cm,keepaspectratio]{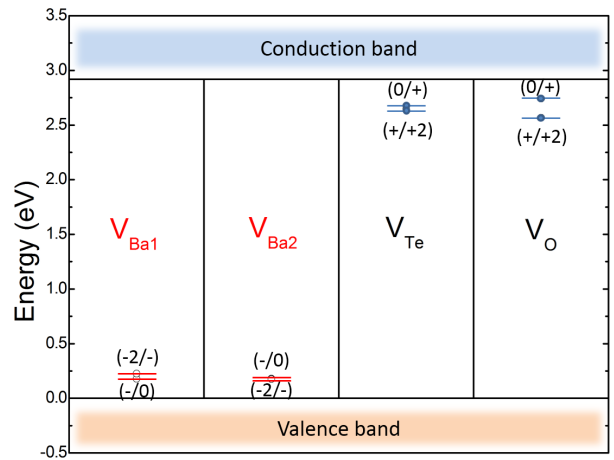}
\caption{\label{fig6}Calculated transition energy levels for various vacancies. The hollow (solid) points show the number of holes (electrons) that can be released between different charge states.}
\end{figure} 

On the other hand, under Ba-poor conditions, the formation energy of V$_{\rm{Ba2}}$ crosses that of V$_{\rm{O}}$, which results in pinning the Fermi level at about 1.1 eV above the VBM, below the middle of the band gap. The formation energy of charged barium vacancies becomes negative when the Fermi level shifts towards the CBM indicating the limitation of n-type doping. Thus, the undoped Ba$_{2}$TeO is expected to be slightly p-type under this condition. 

In addition to the vacancies, the interstitial defects were also considered but no stable positions found. This may due to the fact that Ba$_{2}$TeO is a highly ionic compound with a rather compact structure which constrains the interstitial positions.   

It is also important to investigate the transition energy levels $\epsilon_{\alpha}(q/q^{'})$ that can help in the understanding of the contribution of the defects on the electrical conductivity and the electron-hole recombination effects. In Fig. \ref{fig6}, the calculated transition energy levels from Eq. (\ref{eq8}) are plotted relative to the valence band and conduction band edges. As can be seen, both V$_{\rm{Ba1}}$ and V$_{\rm{Ba2}}$ sites are shallow acceptors with less than 50 meV energy difference between the two charge states. Particularly, energy levels are E(-/0) = E$_{v}$ + 0.18 eV and E(-2/-) = E$_{v}$ + 0.23 eV, for V$_{\rm{Ba1}}$. V$_{\rm{Ba2}}$ even has a lower energy of (-2/-) instead of (-/0), which can be seen in the figure. For the charged V$_{\rm{Te}}$, the calculated energy levels are: E(+/0) = E$_{c}$ - 0.24 eV and E(+2/+) = E$_{c}$ - 0.27 eV. For V$_{\rm{O}}$, one shallow level occur at 0.18 eV and a relative deep level at 0.35 eV below CBM are found.

\section{\label{sec:level4}CONCLUSIONS}
In conclusion, we have studied the band structure, optical properties, and native defects for the new semiconducting material Ba$_{2}$TeO. The band gap of Ba$_{2}$TeO is only slightly below the ideal for a TCM ($>$ 3.1 eV) and we find light electrons and holes ($<$ 1 m$_{e}$) in both the $\Gamma$-X and $\Gamma$-Z directions. However, the defect analysis suggests a spontaneous formation of donor defects including oxygen vacancies and tellurium vacancies in both Ba-rich and Ba-poor conditions. This is detrimental to the p-type transport property of the material since the Fermi level will not be pined close to VBM which will further constrain the concentration and conductivity of the holes. Nevertheless, the absorption features in the infrared-red region suggesting potential energy filter applications provided it can be doped. Finding ways of doping this material will be the critical challenge for applications.

\begin{acknowledgments}
Work at ORNL was supported by the Department of Energy, Basic Energy Science, Materials Sciences and Engineering Division. JS acknowledges a graduate student fellowship, funded by the Department of Energy, Basic Energy Science, Materials Sciences and Engineering Division, through the ORNL GO! program. A portion of this work was performed at the high-performance computing center at the National High Magnetic Field Laboratory.
\end{acknowledgments}

\bibliography{Ba2TeO}

\end{document}